\documentclass[aps,prd,twocolumn,preprintnumbers,showpacs,nofootinbib,amssymb]{revtex4}
\usepackage{graphicx}
\usepackage{amsmath}
\usepackage{amssymb}
\usepackage{amsfonts}
\usepackage{bm}
\usepackage{color}

\def\be{\begin{equation}}
\def\ee{\end{equation}}
\def\bea{\begin{eqnarray}}
\def\eea{\end{eqnarray}}

\begin{document}

\title{New predictions from the logotropic model}
\author{Pierre-Henri Chavanis}
\affiliation{Laboratoire de
Physique Th\'eorique, Universit\'e de Toulouse, CNRS, UPS, France }

\begin{abstract} 

In a previous paper [P.H. Chavanis, Eur. Phys. J. Plus {\bf 130}, 130 (2015)]
we have introduced a new cosmological model that we called
the logotropic model. This model involves a 
fundamental constant $\Lambda$ which is the counterpart of Einstein's
cosmological
constant in the $\Lambda$CDM model. The logotropic model is able to account,
without free parameter, for the constant surface density of the dark matter
halos, for their mass-radius
relation, and for the Tully-Fisher relation. In this paper, we explore other
consequences of this model. By advocating a form of ``strong cosmic
coincidence'' we predict that the present proportion of dark
energy in the Universe is
$\Omega_{\rm de,0}=e/(1+e)\simeq 0.731$ which is close to the
observed value. We also remark that the surface density of dark matter halos and
the surface density of
the Universe are of the same order as
the surface density of the electron.
This makes a curious connection between cosmological and atomic scales. Using
these coincidences, we can relate the Hubble constant, the electron mass and the
electron charge to the cosmological
constant. We also suggest that the famous numbers $137$ (fine-structure
constant) and $123$ (logotropic constant) may actually represent
the same thing. This could unify microphysics and cosmophysics. We study the
thermodynamics of the logotropic model
and find a connection to the Bekenstein-Hawking  entropy of black holes if we
assume that the logotropic
fluid is made of particles of mass $m_{\Lambda}\sim
\hbar\sqrt{\Lambda}/c^2=2.08\times 10^{-33}\,
{\rm eV/c^2}$ (cosmons). In that case, the universality of the surface density
of the dark matter halos  may be related to
a form of holographic principle (the fact that their entropy scales like their
area). We use similar arguments to explain why the surface density of the
electron and the surface density of the Universe are of the same order and
justify the
empirical Weinberg relation. Finally, we combine the
results of our approach with the quantum
Jeans instability theory to predict the order of magnitude of the
mass of
ultralight axions $m\sim 10^{-23}\,
{\rm eV/c^2}$ in the Bose-Einstein condensate
dark matter paradigm. 

\end{abstract}


\maketitle

\section{Introduction}

The nature of dark matter and dark energy remains one of the greatest
mysteries of modern cosmology. Dark matter is responsible for the flat rotation
curves of the galaxies and dark energy is responsible for the accelerated
expansion of the Universe. It is found that dark energy represents about $70\%$
of the energy content of the present Universe while the proportions of dark
matter and baryonic matter are $25\%$ and $5\%$ respectively.

In a previous paper \cite{epjp} (see also \cite{lettre,jcap}) we have
introduced a new cosmological
model that we called
the logotropic model. In this model, there is no dark matter and no dark energy.
There is just a single dark fluid. What we call ``dark matter'' actually
corresponds to its rest-mass energy and what we call ``dark energy'' corresponds
to its internal energy.\footnote{Many models try to
unify dark matter and dark energy. They are called  unified dark energy and dark
matter (UDE/M) models. However, the interpretation of dark matter
and dark energy that we give in Refs. \cite{epjp,lettre} is new and
original.}

Our model does not contain any arbitrary parameter so that it is totally
constrained. It involves a fundamental constant $\Lambda$ which is the
counterpart of Einstein's cosmological constant \cite{einsteincosmo} in the
$\Lambda$CDM (cold dark
matter) model and which turns out to have the same value. Still the
logotropic model
is
fundamentally different from the $\Lambda$CDM model.

On the large (cosmological) scales, the   logotropic model is indistinguishable
from the $\Lambda$CDM model up to the present epoch \cite{epjp,lettre,jcap}. The
two
models
will differ in the far future, in about $25\, {\rm Gyrs}$ years, after which the
 logotropic model will become phantom (the
energy density will increase as
the Universe expands)
and present a Little Rip (the
energy density and the scale factor will become infinite in
infinite time) contrary to the $\Lambda$CDM
model in which the energy density tends towards a  constant 
(de Sitter era).

On the small (galactic) scales, the logotropic model is able to solve
some of the problems encountered by the $\Lambda$CDM model \cite{epjp,lettre}.
In
particular, it is able to account, without free parameter, for the
constant surface density of the dark matter halos, for their mass-radius
relation, and for the Tully-Fisher relation. 

In this paper, we explore other consequences of this model. By advocating a form
of ``strong cosmic coincidence'', stating that the present value of
the dark energy density $\rho_{\rm de, 0}$ is equal to the fundamental constant
$\rho_\Lambda$ appearing in the logotropic model, we predict that
the present proportion of dark energy in the Universe is $\Omega_{\rm
de,0}=e/(1+e)=0.731$ which is close to the observed value
$0.691$ \cite{planck2016}. The
consequences of this result, which implies that our epoch is very special in the
history of the Universe, are intriguing and related to a form of
 anthropic cosmological
principle \cite{barrow}.

We also remark that the universal surface density of dark matter halos (found
from the observations \cite{donato} and predicted by our model
\cite{epjp,lettre}) and the surface density of
the Universe are of the same order of magnitude as the surface density of the
electron.
This makes a curious connection between cosmological and atomic scales.
Exploiting this coincidence, we can relate the Hubble constant, the
electron mass and the electron charge to the
cosmological constant $\Lambda$.  We also argue that the famous
numbers $137$ (fine-structure constant) and $123$ (logotropic
constant) may actually represents
the same thing. This may be a hint for a theory of
unification of microphysics and cosmophysics. Speculations are made in the
Appendices to try to relate these  interconnections to a form of holographic
principle \cite{bousso} stating that the entropy of the electron, the
entropy of dark matter
halos, and the entropy of the Universe scales like their area as in the case of
the entropy of black
holes \cite{bekenstein,hawking}.

\section{The logotropic model}
\label{sec_lm}

\subsection{Unification of dark matter and dark energy}

The Friedmann equations for a flat universe without cosmological constant
are \cite{weinbergbook}:
\begin{equation}
\frac{d\epsilon}{dt}+3\frac{\dot a}{a}(\epsilon+P)=0,\quad H^2=\left
(\frac{\dot a}{a}\right )^2=\frac{8\pi
G}{3c^2}\epsilon,
\label{lm1}
\end{equation}
where $\epsilon(t)$ is the energy density of the Universe, $P(t)$ is the
pressure, $a(t)$ is the
scale factor, and $H=\dot a/a$ is the Hubble parameter.

For a relativistic fluid experiencing an
adiabatic evolution such that $Td(s/\rho)=0$, the first law of
thermodynamics
reduces to \cite{weinbergbook}:
\begin{equation}
d\epsilon=\frac{P+\epsilon}{\rho}d\rho,
\label{lm2}
\end{equation}
where $\rho$ is the rest-mass density of the Universe. Combined
with the equation of continuity
(\ref{lm1}), we get
\begin{equation}
\frac{d\rho}{dt}+3\frac{\dot a}{a}\rho=0 \Rightarrow \rho=\frac{\rho_0}{a^3},
\label{lm3}
\end{equation}
where  $\rho_0$ is the present value of the rest-mass density (the
present
value of the scale factor is taken to be $a=1$). This equation, which
expresses
the conservation of the rest-mass, is valid
for an arbitrary
equation
of state.

For an equation of state specified under the form  $P=P(\rho)$, Eq.
(\ref{lm2}) can be integrated to obtain the relation between the energy density
$\epsilon$ and the rest-mass density. We obtain \cite{epjp}:
\begin{equation}
\epsilon=\rho c^2+\rho\int^{\rho}\frac{P(\rho')}{{\rho'}^2}\, d\rho'=\rho
c^2+u(\rho).
\label{lm4}
\end{equation}
We note that $u(\rho)$
can be interpreted as an internal energy density \cite{epjp}. Therefore, the
energy density $\epsilon$ is the sum of the rest-mass energy $\rho c^2$ and the
internal energy $u(\rho)$.

\subsection{The logotropic dark fluid}
\label{sec_ldf}

We assume that the Universe is filled with a single dark fluid described by
the
logotropic equation of state \cite{epjp}:
\begin{equation}
P=A\ln\left (\frac{\rho}{\rho_P}\right ),
\label{lm5}
\end{equation}
where $\rho_P=c^5/\hbar G^2=5.16\times 10^{99}\, {\rm g\, m^{-3}}$ is the Planck
density and $A$ is a new fundamental constant
of physics, with the dimension of an energy density, which is the
counterpart of the cosmological constant $\Lambda$ in the $\Lambda$CDM model
(see below).
Using Eqs. (\ref{lm4}) and (\ref{lm5}), the relation
between the energy density and the rest-mass density is
\begin{equation}
\epsilon=\rho c^2-A\ln \left (\frac{\rho}{\rho_P}\right )-A=\rho c^2+u(\rho).
\label{lm6}
\end{equation}
The energy density is the
sum of two
terms: a rest-mass energy term $\rho c^2=\rho_0c^2/ a^{3}$ that mimics the
energy density $\epsilon_{\rm m}$ of dark matter and
an internal energy term $u(\rho)=-A\ln \left ({\rho}/{\rho_P}\right )-A
=-P(\rho)-A=3A\ln a-A\ln(\rho_0/\rho_P)-A$
that mimics the
energy density $\epsilon_{\rm de}$ of dark energy. This
decomposition
leads to a natural, and physical, unification of dark matter and dark energy and
elucidates their
mysterious nature. 

Since, in our model, the rest-mass energy of the dark fluid  mimics dark matter,
we
identify $\rho_0c^2$ with the present energy density of dark matter. We thus
set $\rho_0c^2=\Omega_{\rm m,0}\epsilon_0$,
where $\epsilon_0/c^2={3H_0^2}/{8\pi G}$ is the
present energy density of the Universe  and 
$\Omega_{\rm m,0}$ is the present fraction of dark matter (we also include
baryonic
matter). As a result, the present internal energy of the dark
fluid,  $u_0=\epsilon_0-\rho_0c^2$, is identified with the present
dark energy density $\epsilon_{\rm de,0}=\Omega_{\rm de,0}\epsilon_0$
where
$\Omega_{\rm de,0}=1-\Omega_{\rm m,0}$ is the present
fraction of dark energy. Applying Eq. (\ref{lm6}) at the present epoch ($a=1$),
we
obtain the identity
\begin{equation}
A=\frac{\epsilon_{\rm de,0}}{\ln\left(\frac{\rho_Pc^2}{\epsilon_{\rm
de,0}}\right
)+\ln\left (\frac{\Omega_{\rm de,0}}{1-\Omega_{\rm de,0}}\right )-1}.
\label{lm7}
\end{equation}
At that stage, we can have two points of view. We can consider that this
equation determines the constant $A$ as a function of
$\epsilon_{0}$ and $\Omega_{\rm de,0}$ that are both obtained from the
observations \cite{planck2016}. This allows us to determine the value of $A$.
This is the
point of view that we have adopted in our previous papers \cite{epjp,lettre} and
that we adopt in Sec. \ref{sec_B} below. However, in the following section,
we present another point of view leading to an intriguing result.

\subsection{Strong cosmic coincidence and
prediction of $\Omega_{\rm de,0}$}

Let us recall that, in our model, $A$ is considered as a fundamental constant
whose value is fixed by Nature. As a result, Eq. (\ref{lm7}) relates
$\Omega_{\rm de,0}$ to $\epsilon_{0}$ for a given value of $A$. A priori, we
have two unknowns for just one equation. However, we can
obtain the value of $\Omega_{\rm de,0}$ by the following argument.

We can always write the constant  $A$ under the form
\begin{equation}
A=\frac{\rho_{\Lambda}c^2}{\ln\left(\frac{\rho_P}{\rho_{\Lambda}}\right
)}.
\label{lm8}
\end{equation}
This is just a change of notation. Eq. (\ref{lm8}) defines a new
constant, the cosmological
density $\rho_{\Lambda}$, in place of $A$.  From the cosmological density
$\rho_{\Lambda}$,
we can define an effective cosmological constant $\Lambda$ by\footnote{We stress
that our
model is different from the $\Lambda$CDM model so that $\Lambda$ is
fundamentally different from Einstein's cosmological constant
\cite{einsteincosmo}. However, it is always possible to  introduce from the
constant $A$ an effective
cosmological density $\rho_{\Lambda}$ and an effective cosmological constant
$\Lambda$ by Eqs. (\ref{lm8}) and (\ref{lm9}).}
\begin{equation}
\rho_{\Lambda}=\frac{\Lambda}{8\pi G}.
\label{lm9}
\end{equation}
Again this is just a change of notation. Therefore, the fundamental constant
of our model is either $A$, $\rho_{\Lambda}$ or $\Lambda$ (equivalently). We now
advocate a form of ``strong cosmic coincidence''. We assume that the present
value of the dark energy density is equal to $\rho_{\Lambda}c^2$, i.e.,
\begin{equation}
\epsilon_{\rm de,0}=\rho_{\Lambda}c^2.
\label{lm10}
\end{equation}
Since, in the $\Lambda$CDM model, $\epsilon_{\rm de}$ is a constant usually
measured at the present epoch our
postulate implies that $\rho_{\Lambda}c^2$ coincides with the
cosmological density in the $\Lambda$CDM model and that $\Lambda$, as defined
by Eq. (\ref{lm9}), coincides with
the ordinary cosmological constant. This is why we have used the same
notations. Now, comparing Eqs. (\ref{lm7}), (\ref{lm8}) and  (\ref{lm10})  we
obtain $\ln\left
\lbrack \Omega_{\rm de,0}/(1-\Omega_{\rm de,0})\right \rbrack-1=0$
which determines $\Omega_{\rm de,0}$. We find that
\begin{equation}
\Omega_{\rm de,0}^{\rm th}=\frac{e}{1+e}\simeq 0.731
\label{lm11}
\end{equation}
which is close to the observed value $\Omega_{\rm de,0}^{\rm
obs}=0.691$ \cite{planck2016}.  This agreement is puzzling. It
relies on the ``strong cosmic
coincidence'' of Eq. (\ref{lm10}) implying that our epoch is very special. This
is a form of  anthropic cosmological principle \cite{barrow}. This may also
correspond to a fixed point of our model. In order to avoid
philosophical issues, in the following, we adopt the more conventional
point of view discussed at the end of Sec. \ref{sec_ldf}.

\subsection{The logotropic constant $B$}
\label{sec_B}

We can rewrite Eq. (\ref{lm8}) as
\begin{equation}
A=B\rho_{\Lambda}c^2\qquad {\rm
with}\qquad B=\frac{1}{\ln\left({\rho_P}/{\rho_{\Lambda}}
\right
)}.
\label{lm12}
\end{equation}
Again, this is just a change of notation defining the dimensionless number
$B$. We shall call it the logotropic constant since it is equal
to the inverse of the logarithm of the cosmological density normalized by the
Planck density (see Appendix \ref{sec_const}).
We note that $A$ can be expressed in terms of $B$ (see below) so that the
fundamental constant of our model is either $A$, $\rho_{\Lambda}$, $\Lambda$,
or $B$. In the following, we shall express all the results in terms of $B$.
For example, the relation (\ref{lm6}) between the energy density and the scale
factor can be rewritten as 
\begin{equation}
\frac{\epsilon}{\epsilon_0}=\frac{\Omega_{\rm
m,0}}{a^3}+(1-\Omega_{\rm m,0})(1+3B\ln
a).
\label{lm13}
\end{equation}
Combined with the Friedmann equation (\ref{lm1}) this equation determines the
evolution of the scale factor $a(t)$ of the Universe in the logotropic model.
This evolution has been studied in detail
in  \cite{epjp,lettre,jcap}.

{\it Remark:} Considering Eq. (\ref{lm13}), we see that the
$\Lambda$CDM model is recovered for $B=0$.
According to Eq. (\ref{lm12}) this implies that
$\rho_P\rightarrow +\infty$, i.e., $\hbar\rightarrow 0$. Therefore, the 
$\Lambda$CDM model corresponds to the semiclassical limit of the logotropic
model. The fact that $B$ is intrinsically nonzero implies that
quantum mechanics ($\hbar\neq 0$) plays some role in our model in addition to
general relativity. This may
suggest a link with a theory of quantum gravity.

\subsection{The value of $B$ from the
observations}
\label{sec_Bobs}

The fundamental constant ($A$, $\rho_{\Lambda}$, $\Lambda$,
or $B$) appearing in our model can be  determined from the
observations by using Eq. (\ref{lm7}). We take $\Omega_{\rm de,0}=0.6911$ and
 $H_0=2.195\times 10^{-18}\, {\rm
s^{-1}}$ \cite{planck2016}. This implies $\epsilon_0/c^2=3H_0^2/8\pi
G=8.62\times 10^{-24}\, {\rm g\, m^{-3}}$ and $\epsilon_{\rm
de,0}/c^2=\Omega_{\rm
de,0}\epsilon_0/c^2=5.96\times 10^{-24}\, {\rm g\, m^{-3}}$. Since $\ln\left
\lbrack \Omega_{\rm de,0}/(1-\Omega_{\rm de,0})\right \rbrack-1=-0.195$
is small as
compared to $\ln(\rho_Pc^2/\epsilon_{\rm de,0})=283$, we can write in very
good approximation $A$ as in Eq. (\ref{lm8}) with $\rho_{\Lambda}\simeq
\epsilon_{\rm de,0}/c^2$ as in Eq. (\ref{lm10}). Therefore, 
\begin{equation}
\rho_{\Lambda}=\frac{3\Omega_{\rm
de,0}H_0^2}{8\pi G}=5.96\times 10^{-24}\, {\rm g\, m^{-3}}
\label{lm14a}
\end{equation}
and
\begin{equation}
\Lambda= 3\Omega_{\rm
de,0}H_0^2=1.00\times 10^{-35}\, {\rm s^{-2}}
\label{lm14b}
\end{equation}
are approximately equal to
the cosmological density and to the cosmological constant in the $\Lambda$CDM
model. From Eq. (\ref{lm12}) we get
\begin{equation}
B=\frac{1}{\ln(\rho_P/\rho_{\Lambda})}\simeq
\frac{1}{123\ln(10)}\simeq 3.53\times 10^{-3}.
\label{lm15}
\end{equation}
As discussed in our previous papers \cite{epjp,lettre,jcap}, $B$ is
essentially the inverse of
the
famous number $123$ (see Appendix \ref{sec_const}). Finally,
\begin{equation}
A=B\,\rho_{\Lambda}c^2=1.89\times 10^{-9}
\, {\rm g}\, {\rm m}^{-1}\, {\rm s}^{-2}. 
\label{lm16}
\end{equation}

From now on, we shall view $B$ given by Eq. (\ref{lm15}) as the fundamental
constant of the theory. Therefore, everything should be expressed in terms of
$B$ and the other fundamental constants of physics defining the Planck scales.
First, we have
\begin{equation}
\frac{\rho_{\Lambda}}{\rho_{P}}=\frac{G\hbar\Lambda}{8\pi
c^5}=e^{-1/B}=1.16\times 10^{-123}.
\label{lm18}
\end{equation}
Then,
\begin{equation}
\frac{A}{\rho_{P}c^2}=Be^{-1/B}=4.08\times 10^{-126}.
\label{lm17}
\end{equation}
The logotropic equation of state (\ref{lm5}) can be written as
$P/\rho_Pc^2=Be^{-1/B}\ln(\rho/\rho_P)$. Using Eq. (\ref{lm10}) and
$\epsilon_{\rm
de,0}=\Omega_{\rm de,0}\epsilon_0$, we get
\begin{equation}
\frac{\epsilon_0}{\rho_{P}c^2}=\frac{1}{\Omega_{\rm de,0}}e^{-1/B}=1.67\times
10^{-123}.
\label{lm19}
\end{equation}
Finally, using Eq. (\ref{lm1}),
\begin{equation}
t_P H_0=\left (\frac{8\pi}{3\Omega_{\rm de,0}}\right
)^{1/2}e^{-1/2B}=1.18\times 10^{-61},
\label{lm20}
\end{equation}
where $t_P=(\hbar G/c^5)^{1/2}=5.391\times 10^{-44}\, {\rm s}$ is the Planck
time. In the last two expressions,
we can either consider that $\Omega_{\rm de,0}$ is ``predicted'' by
Eq. (\ref{lm11}) or take its measured value. To the order of
accuracy that we
consider, this does not change the numerical values.

\section{Previous predictions of the logotropic model}

The interest of the logotropic model becomes apparent when it is
applied to dark matter halos \cite{epjp,lettre}. We assume that dark matter
halos are
described by the logotropic equation of state of Eq. (\ref{lm5}) with
$A=1.89\times 10^{-9} \, {\rm g}\, {\rm m}^{-1}\, {\rm
s}^{-2}$ (or $B=3.53\times 10^{-3}$). At the
galactic scale, we can use Newtonian gravity.

\subsection{Surface density of dark matter halos}

It is an empirical evidence that the surface density of galaxies has
the same value 
\begin{equation}
\Sigma_0^{\rm obs}\equiv \rho_0 r_h\simeq 295\, {\rm g\, m^{-2}}\simeq 141\,
M_{\odot}/{\rm pc^2}
\label{lm21}
\end{equation}
even
if their sizes and masses vary by several orders of magnitude (up to $14$
orders of magnitude in luminosity) \cite{donato}. Here $\rho_0$ is the central
density and $r_h$ is the halo radius at which the density has decreased by a
factor of $4$. The logotropic model predicts that the surface density of the
dark matter halos is the
same for all the halos (because $A$ is a universal constant) and that it  is
given by \cite{epjp,lettre}:
\begin{equation}
\Sigma_0^{\rm th}=\left (\frac{A}{4\pi G}\right )^{1/2}\xi_h= \left
(\frac{B}{32}\right
)^{1/2}\frac{\xi_h}{\pi}\frac{c\sqrt{\Lambda}}{G},
\label{lm22}
\end{equation}
where $\xi_h=5.8458...$ is a pure number arising from the Lane-Emden equation
of index $n=-1$ expressing the condition of hydrostatic equilibrium of
logotropic
spheres.\footnote{The logotropic spheres \cite{epjp,lettre}, like the isothermal
spheres \cite{chandra}, have an infinite mass. This implies that the logotropic
equation of state cannot
describe dark matter halos at infinitely large distances. Nevertheless, it may
describe the inner region of dark matter halos and this is sufficient to
determine their surface density. The stability of bounded logotropic spheres has
been studied in \cite{logo} by analogy with the stability of bounded isothermal
spheres and similar results have been obtained. In particular, bounded
logotropic
spheres are stable
provided that the density contrast is not too large.} Numerically,
\begin{equation}
\Sigma_0^{\rm th}= 278\, {\rm g\,
m^{-2}}\simeq 133\,
M_{\odot}/{\rm pc^2},
\label{lm23}
\end{equation}
which is very close to the observational value
(\ref{lm21}). The fact that the surface density of dark matter halos is
determined by the effective cosmological constant $\Lambda$ (usually related to
the dark energy) tends to confirm that dark matter and dark energy are just two
manifestations of the {\it same} dark fluid, as we have
assumed in our model.

{\it Remark:} The dimensional term $c\sqrt{\Lambda}/G$ in Eq. (\ref{lm22}) can
be interpreted as representing the surface density of the Universe (see Appendix
\ref{sec_w}).  We note that this term alone, $c\sqrt{\Lambda}/G=14200\, {\rm g\,
m^{-2}}=6800 M_{\odot}/{\rm pc}^2$, is too large to account precisely for the
surface density of dark matter halos so that the prefactor
$(B/32)^{1/2}(\xi_h/\pi)=0.01955$ is
necessary to reduce this number.  It is interesting to remark
that the term $c\sqrt{\Lambda}/G$ arises from classical general relativity while
the prefactor $\propto B^{1/2}$ has a quantum origin as discussed at the end of
Sec. \ref{sec_B}. Actually, we will see that it is related to the fine-structure
constant $\alpha$ [see Eq. (\ref{lm30}) below].

\subsection{Mass-radius relation}

There are interesting consequences of the preceding result. For logotropic
halos, the mass of the halos calculated at the halo
radius $r_h$ is given by  \cite{epjp,lettre}:
\begin{equation}
M_h=1.49\Sigma_0 r_h^2.
\label{lm24a}
\end{equation}
This
determines the mass-radius relation of dark matter-halos. On the other hand, the
circular
velocity at the halo radius is $v_h^2=GM_h/r_h=1.49\Sigma_0
G r_h$. Since the surface density of
the dark matter halos is constant, we obtain
\begin{equation}
\frac{M_h}{M_{\odot}}=198 \left (\frac{r_h}{{\rm pc}}\right )^2,\qquad
\left (\frac{v_h}{{\rm
km}\, {\rm s}^{-1}}\right )^2=0.852\, \frac{r_h}{\rm pc}.
\label{lm24}
\end{equation}
The scalings $M_h\propto r_h^2$ and $v_h^2\propto
r_h$ (and also the prefactors) are consistent with the observations.

\subsection{The Tully-Fisher relation}

Combining the previous equations, the logotropic model leads to the Tully-Fisher
\cite{tf} relation $v_h^4\propto M_h$ or, more precisely, 
\begin{equation}
\left (\frac{M_b}{v_h^4}\right )^{\rm th}=\frac{f_b}{1.49\Sigma_0^{\rm th}
G^2}=46.4 M_{\odot}{\rm km}^{-4}{\rm s}^4,
\label{lm25}
\end{equation}
where
$f_b=M_b/M_h\sim 0.17$ is the cosmic baryon fraction \cite{mcgaugh}. The
predicted value from Eq. (\ref{lm25}) is
close to the observed one $\left ({M_b}/{v_h^4}\right )^{\rm obs}=47\pm 6
M_{\odot}{\rm
km}^{-4}{\rm
s}^4$ \cite{mcgaugh}.

{\it Remark:} The Tully-Fisher relation is sometimes justified by
the MOND (Modification of Newtonian dynamics) theory \cite{mond} which predicts
a relation of the form $v_h^4=Ga_0
M_b$ between the asymptotic circular velocity and the baryon mass, where $a_0$
is a critical acceleration. Our results imply
$a_0^{\rm th}=1.62\times
10^{-10}\, {\rm m}\, {\rm s}^{-2}$ which is
close to the value $a_0^{\rm obs}=(1.3\pm 0.3)\times
10^{-10}\, {\rm m}\, {\rm s}^{-2}$ obtained from the observations
\cite{mcgaugh}.
Combining Eqs. (\ref{lm24a}) and (\ref{lm25}), we first get $a_0^{\rm
th}=(1.49/f_b)\Sigma_0^{\rm th}G=GM_h/(f_br_h^2)$ which shows that $a_0$ can be
interpreted as the surface gravity of the galaxies $G\Sigma_0$ (which
corresponds to Newton's acceleration $GM_h/r_h^2$) or as the surface density of
the Universe (see Appendix \ref{sec_abh}). Then, using Eqs.
(\ref{lm14b}) and (\ref{lm22}), we obtain $a_0^{\rm
th}=({1.49}/{f_b})({B}/{32})^{1/2}({\xi_h}/{\pi})c\sqrt{\Lambda}\simeq H_0
c/4$  which
explains why $a_0$ is of the order of $H_0 c$. We emphasize,
however, that we do
not use the MOND theory in our approach and that the logotropic model assumes
the existence of a dark fluid.

\subsection{The mass $M_{300}$}

The logotropic equation of state also explains the observation of Strigari {\it
et al.} \cite{strigari} that all the dwarf spheroidals (dSphs) of the Milky
Way have the same total dark matter mass $M_{300}$ contained within a radius
$r_u=300\, {\rm pc}$, namely $M_{300}^{\rm obs}\simeq 10^7\, M_{\odot}$
The logotropic model predicts the
value \cite{epjp,lettre}:
\begin{equation}
M_{300}^{\rm th}=\frac{4\pi \Sigma_0^{\rm th} r_u^2}{\xi_h\sqrt{2}}=1.82\times
10^{7}\, M_{\odot},
\label{lm26}
\end{equation}
which is in very good agreement with the
observational value.

\section{A curious connection between atomic and
cosmological scales}
\label{sec_curious}

\subsection{The surface density of the electron}
\label{sec_sde}

The classical radius of the
electron $r_e$ can be obtained qualitatively by writing that the electrostatic
energy of the electron, $e^2/r_e$, is equal to its rest-mass energy $m_e c^2$.
Recalling the value of the charge of the electron $e=4.80\times 10^{-13}\, {\rm
g^{1/2}\, m^{3/2}\, s^{-1}}$ and its mass $m_e=9.11\times 10^{-28}\, {\rm g}$,
we obtain $r_e=e^2/m_ec^2=2.82\times 10^{-15}\, {\rm m}$. As a result, the
surface density of the electron is\footnote{We note that the Thomson
cross-section $\sigma=(8\pi/3)(e^2/m_e
c^2)^2$ can be written as $\sigma=(8\pi/3)r_e^2$ giving a physical meaning to
the classical electron radius $r_e$. We also note that $r_e$ can be written as 
$r_e=\alpha\hbar/m_ec$ where $\lambda_C=\hbar/m_e c$ is the Compton
wavelength of the electron and $\alpha$ is the fine-structure constant $\alpha$
[see Eq. (\ref{const1})]. Similarly, 
$\Sigma_e=(1/\alpha^2)m_e^3c^2/\hbar^2$.}
\begin{equation}
\Sigma_e=\frac{m_e}{r_e^2}=\frac{m_e^3c^4}{e^4}=115\, {\rm g/m^2}= 54.9\,
M_{\odot}/{\rm pc^2},
\label{lm27}
\end{equation}
which is of the same order of magnitude as the surface density of dark matter
halos from Eq. (\ref{lm21}). This
coincidence is amazing in view of the different scales (atomic
versus cosmological) involved.  More precisely, we find
$\Sigma_e=\sigma\Sigma_0^{\rm th}$ with $\sigma\simeq 0.413$. Of
course, the value of $\sigma$ depends on the
precise manner used to define the surface density of the electron, or its
radius, but the
important point is that this number is of order unity.

\subsection{Relation between $\alpha$ and $B$}
\label{sec_Ba}

By matching the two formulae (\ref{lm22}) and (\ref{lm27}),
writing $\Sigma_e=\sigma\Sigma_0^{\rm th}$, we get
\begin{equation}
\Lambda=\frac{32\pi^2}{B\xi_h^2\sigma^2}
\frac{m_e^6c^6G^2}{e^8}=\frac{32\pi^2}{B\xi_h^2\sigma^2\alpha^4}
\frac{m_e^6c^2G^2}{\hbar^4},
\label{lm28}
\end{equation}
where we have introduced the fine-structure constant $\alpha$ in the
second equality (see Appendix \ref{sec_const}).
This expression provides a curious relation between the cosmological constant,
the mass of the electron and its charge. This relation is similar to
Weinberg's empirical relation (see Appendix
\ref{sec_w}) which can be written as [combining Eqs. (\ref{lm14b}) and
(\ref{w4})]
\begin{equation}
\Lambda=192\pi^2\mu^2\Omega_{\rm de,0}\frac{m_e^6c^6G^2}{e^8},
\label{w4b}
\end{equation}
where $\mu\simeq 3.42$. Note that in our formula (\ref{lm28}),
$\Lambda$ appears two times: on the left hand side and in $B$ (which depends
logarithmically on $\Lambda$). This will have important consequences in the
following.

B\"ohmer and Harko \cite{bhLambda}, by a completely different approach, found a
similar relation\footnote{A closely related formula, involving the Hubble
constant instead of the cosmological constant, was first found by Stewart
\cite{stewart} in 1931 by trial and error.}
\begin{equation}
\Lambda=\nu \frac{\hbar^2 G^2 m_e^6 c^8}{e^{12}}=\frac{\nu}{\alpha^6} \frac{G^2
m_e^6 c^2}{\hbar^{4}},
\label{lm29}
\end{equation}
where $\nu\simeq 0.816$ is of order unity. Their result can be
obtained as follows. They first introduce a minimum
mass $m_{\Lambda}\sim\hbar\sqrt{\Lambda}/c^2$ interpreted as being the mass of
the elementary particle of dark energy, called the cosmon. Then, they define a
radius $R$ by the relation $m_{\Lambda}\sim \rho_{\Lambda} R^3$ where
$\rho_{\Lambda}= \Lambda/8\pi G$ is the cosmological density considered as being
the lowest density in the Universe. Finally, they remark that $R$ has
typically the same value as the classical radius of the electron 
$r_e=e^2/m_ec^2$. Matching $R$ and $r_e$ leads to the scaling of Eq.
(\ref{lm29}). We have then added a prefactor $\nu$ and adjusted its
value
in order to exactly obtain the measured value of the  cosmological  constant
\cite{planck2016}.
Since  the approach of B\"ohmer and Harko \cite{bhLambda} is essentially
qualitative, and depends on the precise manner used to define the radius of
the
electron, their result can be at best valid up to a constant of order unity.

We would like now to compare the estimates from Eqs. (\ref{lm28}) and
(\ref{lm29}). At that
stage, we can have two points of view. If we consider that comparing the
prefactors is meaningless because our approach can only provide ``rough'' orders
of
magnitude, we  conclude that Eqs. (\ref{lm28}) and
(\ref{lm29}) are
equivalent, and that they are also equivalent to Weinberg's empirical
relation (\ref{w4}). Alternatively, if we take the prefactors seriously into
account (in particular the presence of $B$ which depends on $\Lambda$) and match
the formulae (\ref{lm28}) and
(\ref{lm29}), we find an interesting relation between the
fine-structure constant
$\alpha$ and the logotropic constant $B$:
\begin{equation}
\alpha=\left (\frac{\nu}{32}\right
)^{1/2}\frac{\xi_h\sigma}{\pi}\sqrt{B}\simeq 0.123 \sqrt{B}.
\label{lm30}
\end{equation}
Therefore, the fine-structure constant (electron charge normalized by the
Planck charge) is determined by the logotropic  constant $B$
(cosmological density normalized by the Planck density) by a relation of the
form $\alpha\propto B^{1/2}$. This makes a connection between atomic scales and
cosmological scales. This also suggests that the famous numbers
$137$ and $123$
(see Appendix \ref{sec_const}) are related to each other, or may even represent
the same thing. From Eq. (\ref{lm30}), we have\footnote{We note that the
prefactors in
Eqs. (\ref{lm30}) and (\ref{lm31}) appear to be close to $123/1000$ and
$123/10$, where the number $123$ appears again (!). We do not know whether this
is
fortuitous or if this bears a deeper significance than is apparent at first
sight.} 
\begin{equation}
137\simeq 12.3 \sqrt{123}.
\label{lm31}
\end{equation}

{\it Remark:} the logotropic constant $B$ is related to the effective
cosmological constant $\Lambda$ by [see Eq.
(\ref{lm18})]
\begin{equation}
B=\frac{1}{\ln\left (\frac{8\pi c^5}{G\hbar\Lambda}\right )}.
\label{lm31b}
\end{equation}
Using Eqs. (\ref{lm30}) and (\ref{lm31b}), we can express the fine-structure
constant $\alpha$ as a function of the effective cosmological
constant $\Lambda$ or, using Eq. (\ref{lm20}), as a function of the age of the
Universe $t_{\Lambda}=1/H_0$ as
\begin{equation}
\alpha=\frac{0.123}{\ln\left (\frac{8\pi c^5}{G\hbar\Lambda}\right
)^{1/2}}=\frac{0.123}{\sqrt{2}\ln\left \lbrack \left (\frac{8\pi}{3\Omega_{\rm
de,0}}\right )^{1/2}\frac{t_{\Lambda}}{t_P}\right\rbrack^{1/2}}.
\label{lm31c}
\end{equation}
We emphasize the scaling $1/\alpha\propto (\ln t_{\Lambda})^{1/2}$. It is
interesting to note that similar relations have been introduced in the past
from pure numerology (see \cite{kragh}, P. 428). These relations suggest
that the fundamental constants may change with time as argued by Dirac
\cite{dirac1,dirac2}.

\subsection{The mass and the charge of the electron
in terms of $B$}

Using Eqs. (\ref{lm9}), (\ref{lm18}), (\ref{lm28}) and (\ref{lm30}), we find
that the mass and the charge of the electron
are determined by the logotropic
constant $B$ according to
\begin{eqnarray}
\frac{m_e}{M_P}=\left (\frac{8\pi}{\nu}\right )^{1/6}\left
(\frac{\nu}{32}\right
)^{1/2}\frac{\xi_h\sigma}{\pi}\sqrt{B}e^{-1/(6B)}\nonumber\\
=0.217\sqrt{B}e^{
-1/(6B) } =4.18\times 10^{-23},
\label{lm32}
\end{eqnarray}
\begin{equation}
\frac{e^2}{q_P^2}=\left (\frac{\nu}{32}\right
)^{1/2}\frac{\xi_h\sigma}{\pi}\sqrt{B}=0.123\sqrt{B}=7.29\times 10^{-3},
\label{lm33}
\end{equation}
where $M_P=(\hbar c/G)^{1/2}=2.18\times 10^{-5}\, {\rm g}$ is the
Planck mass and $q_P=(\hbar c)^{1/2}=5.62\times 10^{-12}\, {\rm
g^{1/2}\, m^{3/2}\, s^{-1}}$ is the Planck charge.
These relations suggest that the  mass and the charge of the electron (atomic
scales) are determined by the effective cosmological constant
$\Lambda$ or $B$ (cosmological scales). We emphasize the presence of
the exponential factor $e^{-1/(6B)}$ in Eq. (\ref{lm32}) explaining why the
electron mass is much smaller than the Planck mass while the electron charge is
comparable to the Planck charge.  

\subsection{A prediction of $B$}

If we match Eqs. 
(\ref{lm22}) and (\ref{w3}), or equivalently Eqs. 
(\ref{lm28}) and (\ref{w4b}), we obtain
\begin{equation}
B^{\rm app}=\frac{1}{6\lambda^2\xi_h^2\Omega_{\rm
de,0}}.
\label{w5}
\end{equation}
Taking $\lambda^{\rm app}=1$ (since we cannot predict its value) and
$\Omega_{\rm de,0}^{\rm th}=e/(1+e)$ [see Eq. (\ref{lm11})], we get $B^{\rm
app}=6.67\times 10^{-3}$ instead of $B=3.53\times
10^{-3}$. We recall
that the value of $B$ was obtained in Sec. \ref{sec_Bobs} from the
observations.
On the other hand, Eq. (\ref{w5}) gives the correct order of magnitude of $B$
without any reference to observations, up to a dimensionless constant
$\lambda\simeq 1.41$ of order unity.
Considering that $B$ is predicted by Eq.
(\ref{w5}) implies that we can predict the values of $\Lambda$, $H_0$, $\alpha$,
$m_e$ and $e$ without reference to observations, up to dimensionless constants
$\lambda\simeq 1.41$, $\nu\simeq 0.816$ and $\sigma\simeq 0.413$
of order unity.  We note, however, that even if these dimensionless
constants ($\lambda$, $\nu$, $\sigma$) are of order unity, their precise values
are of importance since $B$ usually
appears in exponentials like in Eqs. (\ref{lm18}), (\ref{lm20}) and
(\ref{lm32}).

\section{Conclusion}

In this paper, we have developed the logotropic model introduced in
\cite{epjp,lettre}. In this model, dark matter corresponds to the rest mass
energy of a dark fluid and dark energy  corresponds to its internal energy. The
$\Lambda$CDM model may be interpreted as the semiclassical limit
$\hbar\rightarrow 0$ of the logotropic model. We have first recalled that the 
logotropic model is able to predict (without free parameter) the universal value
of the surface density of dark matter halos $\Sigma_0$, their mass-radius
relation
$M_h-r_h$, the Tully-Fisher relation $M_b\sim v_h^4$ and the value of the mass
$M_{300}$ of dSphs. Then, we have argued that it also predicts the value of the
present fraction of dark energy $\Omega_{\rm de,0}$. This arises from a
sort of ``strong cosmic coincidence'' but this could also correspond to a fixed
point of the model. Finally, we have observed that the surface density of the
dark matter halos  $\Sigma_0$ is of the same order as the surface density of the
Universe
 $\Sigma_\Lambda$ and of the same order as the surface density of the electron 
$\Sigma_e$. This makes an
empirical connection between atomic physics and cosmology. From this connection,
we have obtained a relation between the fine-structure constant $\alpha\sim
1/137$ and the logotropic constant $B\sim 1/123$. We have also expressed the
mass $m_e$  and the charge $-e$ of the electron as a function of $B$
(or as a function of the effective
cosmological constant $\Lambda$). Finally, we have obtained a prediction of the
order of magnitude of $B$ independent from the observations. In a sense, our
approach which expresses the mass and the charge of the electron in terms of
the cosmological constant is a continuation of the program initiated by
Eddington \cite{eddington} in his quest for a `{\it Fundamental Theory}' of
the physical world in which the basic interaction strengths and elementary
particle masses would be prediced entirely combinatorically by simple counting
processes \cite{barrow}. In the Appendices, we try to
relate these interconnections to a form of holographic
principle \cite{bousso} (of course not known at the time of Eddington) stating
that the entropy of the electron, of dark matter
halos, and of the Universe scales like their area as in the case of black
holes \cite{bekenstein,hawking}.

This paper has demonstrated that physics  is full of ``magic'' and mysterious
relations that are still not fully understood (one of them being the empirical
Weinberg relation). Hopefully, a contribution of this
paper is to reveal these ``mysteries'' and propose some tracks so as to  induce
further research towards
their elucidation.

\appendix

\section{The constants $\alpha$ and $B$}
\label{sec_const}

There are two famous numbers in physics, $137$ and $123$, which respectively
apply to atomic and cosmological scales.

At the atomic level, the fine-structure constant $\alpha$, also known as
Sommerfeld's
constant, is a dimensionless physical constant characterizing the strength
of the electromagnetic interaction between elementary charged particles. Its
value is
\begin{equation}
\alpha=\frac{e^2}{\hbar c}=\frac{e^2}{q_P^2}\simeq \frac{1}{137}\simeq
7.30\times 10^{-3}.
\label{const1}
\end{equation}
It can be seen as the square of the charge $e=4.80\times 10^{-13}\, {\rm
g^{1/2}\, m^{3/2}\, s^{-1}}$ of the electron
normalized by the Planck charge $q_P=(\hbar c)^{1/2}=5.62\times 10^{-12}\, {\rm
g^{1/2}\, m^{3/2}\, s^{-1}}$.
The quantum theory does not predict its value. The number $1/\alpha\simeq 137$
intrigued a lot of famous
researchers including Eddington, Pauli, Born, Hawking and Feynman among
others \cite{kragh}. Feynman writes
\cite{feynman}: {\it It's one of the greatest damn mysteries of physics: a magic
number that comes to us with no understanding by man. You might say the ``hand
of God'' wrote that number, and ``we don't know how He pushed his pencil.'' }

At the cosmological  level, there is another famous number 
\begin{equation}
B=\frac{1}{\ln(\rho_P/\rho_{\Lambda})}\simeq
\frac{1}{123\ln(10)}\simeq 3.53\times 10^{-3}.
\label{const2}
\end{equation}
It can be seen as the logarithm of the cosmological - or dark energy -
density 
$\rho_{\Lambda}=\Lambda/8\pi G=5.96\times 10^{-24}\, {\rm g\, m^{-3}}$
(where $\Lambda=1.00\times 10^{-35}\, {\rm s^{-2}}$ is the cosmological
constant),
normalized by the Planck density $\rho_P=c^5/\hbar G^2=5.16\times
10^{99}\, {\rm g\, m^{-3}}$. This number 
appeared in connection to  the so-called cosmological constant problem
\cite{weinbergcosmo,paddycosmo}, i.e., the fact that
there is a difference of $123$ orders of magnitude between the Planck density
and the cosmological density ($\rho_P/\rho_{\Lambda}\sim 10^{123}$) interpreted
as the vacuum energy.

We have suggested in this paper that the two dimensionless constants $\alpha$
and $B$, or the two numbers $137$ and $123$, are related to each other [see
Eqs. (\ref{lm30}) and (\ref{lm31})] and that, in some sense, they correspond to
the same thing. If this idea is correct, it would yield a fascinating connection
between atomic and cosmic physics.

\section{Surface density of the Universe, surface density of the
electron and  Weinberg's empirical relation}
\label{sec_w}

Using qualitative arguments, let us determine the surface
density of the Universe. The Hubble time ($\sim$ age of the Universe) is 
$t_{\Lambda}={1}/{H_0}=14.4$ billion years.
The Hubble radius ($\sim$ radius of the visible Universe) is
$R_{\Lambda}=ct_{\Lambda}=c/H_0=1.37\times 10^{26}\, {\rm m}$.
The present density of the Universe is
${\epsilon_0}/{c^2}={3H_0^2}/{8\pi
G}=8.62\times 10^{-24}\, {\rm g\, m^{-3}}$.
The Hubble mass ($\sim$ mass of the Universe) is
$M_{\Lambda}=({4}/{3})\pi ({\epsilon_0}/{c^2})
R_{\Lambda}^3={c^3}/{2GH_0}=9.20\times 10^{55}\,
{\rm g}$. Combining these relations, we find that the surface density
of the Universe is
\begin{equation}
\Sigma_{\Lambda}=\frac{M_{\Lambda}}{4\pi R_{\Lambda}^2}=\frac{cH_0}{8\pi
G}=392\, {\rm g\, m^{-2}}=188\,
M_{\odot}/{\rm pc^2}.
\label{w2}
\end{equation}
It can be written as $\Sigma_{\Lambda}=cH_0/\kappa c^4$ where $\kappa=8\pi
G/c^4$ is Einstein's
gravitational constant (which includes the $8\pi$ factor). Using Eq.
(\ref{lm14b}), we obtain
\begin{equation}
\Sigma_{\Lambda}=\frac{1}{8\pi\sqrt{3\Omega_{\rm
de,0}}}\frac{c\sqrt{\Lambda}}{G}.
\label{w3}
\end{equation}
This relation shows that the surface density of the Universe provides the
correct scale for the surface density of dark matter halos [see Eq.
(\ref{lm22})].
We have $\Sigma_{\Lambda}=\lambda \Sigma_0^{\rm th}$ with
$\lambda\simeq
1.41$.

Therefore, the surface density of the Universe is of the same
order as
the surface density of the dark matter halos which is also of the same order as
the surface density of the electron (as we have previously observed). We have
$\Sigma_{\Lambda}=\mu \Sigma_e$ with $\mu=\lambda/\sigma\simeq 3.42$.
Matching
Eqs. (\ref{lm27}) and (\ref{w2}), we get 
\begin{equation}
m_e=\left (\frac{e^4H_0}{8\pi\mu Gc^3}\right
)^{1/3}.
\label{w4}
\end{equation}
This relation expresses the mass of the electron as a function of its charge
and the Hubble constant. This mysterious relation 
is mentioned in the book of Weinberg \cite{weinbergbook} where it is obtained
from purely
dimensional arguments.\footnote{Weinberg
considers this
relation as ``so far unexplained'' and having ``a real though mysterious
significance''. Similar relations have been obtained in the past by Stewart
\cite{stewart}, Eddington \cite{eddington} and others from purely heuristic
arguments or from dimensional analysis \cite{kragh,barrow,calogero}. Their
goal was to express the mass of the
elementary particles in terms of the fundamental constants of Nature. } He
observes that the term in the
right hand side
of Eq. (\ref{w4}) has the dimension of a mass
and that this mass, $1.37\times 10^{-27}\, {\rm g}$ (with $\mu^{\rm app}=1$), is
of the order of the mass
of the electron.  The fact
that relation (\ref{w4}) expresses the commensurability of the  
surface density of the Universe and the surface density of the electron, as we
observe here,  may
help elucidating its physical meaning (see Appendix \ref{sec_post}).

{\it Remark:} If the dark matter halos resulted
from
the balance between the
gravitational attraction and the repulsion due to the dark energy, they would
have a typical density $M_h/r_h^3\sim \rho_{\Lambda}$. Actually, such an
equilibrium is unstable as is well-known in the case of the Einstein
static Universe.
Therefore, the radius of dark matter halos must satisfy the constraint
$r_h<(M_h/\rho_{\Lambda})^{1/3}$. Now, we have seen that their mass-radius
relation
scales as $M_h\sim (c\sqrt{\Lambda}/G)r_h^2$. We then find that the constraint 
$r_h<(M_h/\rho_{\Lambda})^{1/3}$ is satisfied provided that
$M_h<c^3/G\sqrt{\Lambda}$.  Since the upper bound is of the order of the mass of
the Universe,
$M_{\Lambda}\sim c^3/G\sqrt{\Lambda}$, we conclude that the size of the dark
matter halos
is always much smaller than the critical size $(r_h)_{\rm
crit}=(M_h/\rho_{\Lambda})^{1/3}$ as required for stability reasons.

\section{Analogy with black hole thermodynamics}
\label{sec_abh}

\subsection{Black hole entropy}
\label{sec_abha}

The Bekenstein-Hawking \cite{bekenstein,hawking} entropy of a Schwarzschild
black hole is given by
\begin{equation}
S_{\rm BH}=\frac{1}{4}k_B \frac{A}{l_P^2}=\frac{k_B \pi c^3R^2}{G\hbar},
\label{bh1}
\end{equation}
where $A=4\pi R^2$ is the area of the event horizon of the black hole and
$l_P=(G\hbar/c^3)^{1/2}=1.62\times 10^{−35}\, {\rm m}$ is the Planck length. The
radius of
a Schwarzschild black hole is connected to its mass by
\begin{equation}
R=\frac{2GM}{c^2}.
\label{bh2}
\end{equation}
The Hawking temperature \cite{hawking} of a  Schwarzschild black hole
is
\begin{equation}
k_B T=\frac{\hbar c^3}{8\pi GM}=\frac{\hbar c}{4\pi R}.
\label{bh3}
\end{equation}
The black hole entropy (\ref{bh1}) can be obtained from the  Hawking temperature
(\ref{bh3}) by using  the thermodynamic relation $T^{-1}=dS_{\rm BH}/d(Mc^2)$.
If
we consider a Planck black hole of radius $l_P$
and mass $M_P$, we find that its
temperature is of the order of the Planck temperature  $T_P=M_P
c^2/k_B=1.42\times 10^{32}\, {\rm K}$ and
its entropy $S_{\rm BH}/k_B\sim 1$.

\subsection{Analogy between the Universe and a black hole}
\label{sec_abhb}

Using the results of Appendix \ref{sec_w}, we note that the radius of the
Universe is related to its mass by  
\begin{equation}
R_{\Lambda}=\frac{2GM_{\Lambda}}{c^2}.
\label{bh4}
\end{equation}
This expression coincides with the mass-radius relation (\ref{bh2}) of a
Schwarzschild black
hole. This coincidence  has sometimes led people to say that the Universe is a
black hole, or that we live in a black hole, although this analogy is probably
too naive. Nevertheless, at least on a purely dimensional basis, we can use
the analogy with black holes to define the entropy and the temperature of the
Universe. In this manner, we get a temperature scale (temperature on the
horizon)
\begin{equation}
k_B T_{\Lambda}=\frac{\hbar c}{4\pi R_{\Lambda}}=\frac{\hbar H_0}{4\pi}\sim
\hbar\sqrt{\Lambda}.
\label{bh5}
\end{equation}
Its value is $T_{\Lambda}\sim 2.41\times 10^{-29}\, {\rm K}$. The temperature
can be
written as
\begin{equation}
k_B T_{\Lambda}=\frac{2\hbar a_{\Lambda}}{c},
\label{w3b}
\end{equation}
where
\begin{equation}
a_{\Lambda}=G\Sigma_{\Lambda}=\frac{GM_{\Lambda}}{4\pi
R_{\Lambda}^2}=\frac{c^2}{8\pi R_{\Lambda}}=
\frac{cH_0}{8\pi}\sim c\sqrt{\Lambda}
\label{w3bb}
\end{equation}
is the surface gravity of the Universe (similar relations apply to black holes).
We can also write
\begin{equation}
k_B T_{\Lambda}=m_{\Lambda}c^2,
\label{bh6}
\end{equation}
with
\begin{equation}
m_{\Lambda}\sim \frac{\hbar\sqrt{\Lambda}}{c^2}=2.08\times 10^{-33}\,
{\rm eV/c^2}.
\label{bh7}
\end{equation}
This mass  scale is often interpreted as the
smallest mass of the bosons predicted by string theory \cite{axiverse} or as
the upper bound on the mass of the graviton \cite{graviton}.\footnote{It is
simply
obtained by equating the Compton wavelength of the particle $\lambda_c=\hbar/mc$
with the Hubble radius $R_\Lambda=c/H_0$ (the typical size of the visible
Universe) giving $m_{\Lambda}=\hbar H_0/c^2$. Using Eq. (\ref{lm14b}), we obtain
Eq. (\ref{bh7}).} It can be
contrasted from the mass scale
\begin{equation}
M_{\Lambda}\sim \frac{c^3}{G\sqrt{\Lambda}}=7.16\times 10^{88}\,
{\rm eV/c^2},
\label{bh8}
\end{equation}
which is usually interpreted as the mass of the Universe. Thus $m_{\Lambda}$
and $M_{\Lambda}$ represent fundamental lower and upper mass scales. Their
ratio is
\begin{equation}
\frac{M_{\Lambda}}{m_{\Lambda}}\sim \frac{c^5}{G\hbar\Lambda}\sim
\frac{\rho_P}{\rho_{\Lambda}}\sim e^{1/B}\sim 10^{123},
\label{bh9}
\end{equation}
which exhibits the famous number $123$. On the other hand, our analogy between
the Universe and a black hole leads to an entropy scale (entropy on the Hubble
horizon):
\begin{equation}
S_{\Lambda}=\frac{k_B \pi c^3R_{\Lambda}^2}{G\hbar}=\frac{k_B \pi
c^5}{G\hbar H_0^2}\sim \frac{k_B c^5}{G\hbar \Lambda}.
\label{bh10}
\end{equation}
We note that the entropy of the Universe can be written as 
\begin{equation}
S_{\Lambda}/k_B\sim  \frac{M_{\Lambda}}{m_{\Lambda}}
\sim  \frac{\Sigma_{\Lambda}R_{\Lambda}^2}{m_{\Lambda}}
\sim  e^{1/B}\sim
10^{123}.
\label{bh11}
\end{equation}
This entropy may be identified with the total entropy of the logotropic dark
fluid (see the Appendix of \cite{jcap} and Appendix \ref{sec_logt}). It can be
compared to the entropy of radiation \cite{cosmopoly1}:
\begin{eqnarray}
S_{\rm rad}/k_B&=&\frac{4}{3}\left (\frac{3\Omega_{\rm rad,0}}{8\pi}\right
)^{3/4}\left (\frac{\pi^2}{15}\right
)^{1/4}\frac{1}{(H_0t_P)^{3/2}}\nonumber\\
&=&5.64\times 10^{87},
\label{bh11b}
\end{eqnarray}
obtained by using Eq. (\ref{t2}) with $P_{\rm rad}=\epsilon_{\rm rad}/3$,
$\epsilon_{\rm rad}=\sigma T^4$ with $\sigma=\pi^2k_B^4/15c^3\hbar^3$
(Stefan-Boltzmann constant), $\epsilon_{\rm rad}=\Omega_{\rm
rad,0}\epsilon_0/a^4$ and $\Omega_{\rm rad,0}=9.24\times  10^{-5}$. They differ
by about $36$ orders of magnitude.

{\it Remarks:} We note that $T_{\Lambda}S_{\Lambda}=(1/2)M_{\Lambda}c^2$ so 
the free energy of the Universe is
$F_{\Lambda}=M_{\Lambda}c^2-T_{\Lambda}S_{\Lambda}=(1/2)M_{\Lambda}c^2$. On
the other hand, using Eqs.
(\ref{lm18}) and (\ref{lm32}), we obtain the relations
\begin{equation}
\frac{m_{\Lambda}}{M_P}\sim e^{-1/2B}=3.40\times 10^{-62},
\label{bh12}
\end{equation}
\begin{equation}
\frac{m_e}{m_{\Lambda}}\sim
\sqrt{B} e^{1/3B}=5.66\times 10^{39}.
\label{bh13}
\end{equation}
Since $m_{\Lambda}\sim \rho_{\Lambda}r_e^3$ (see Sec. \ref{sec_Ba}) we have
$m_e/m_\Lambda\sim \rho_e/\rho_\Lambda$. The gravitational radius of the cosmon
is $r_\Lambda=2Gm_{\Lambda}/c^2\sim G\hbar\sqrt{\Lambda}/c^4=2.75\times
10^{-96}\, {\rm m}$.

\subsection{Entropy of logotropic dark matter halos}

Let us define the entropy of a logotropic dark matter halo by 
\begin{equation}
S\sim k_B N\sim k_B
\frac{M}{m_{\Lambda}},
\label{bh14}
\end{equation} 
where $M$ is the halo mass and $m_{\Lambda}$ is the mass of the
hypothetical particle composing the logotropic dark fluid. Using the
mass-radius
relation $M\sim \Sigma_0 R^2$ of a logotropic dark matter halo, where
$\Sigma_0\sim c\sqrt{\Lambda}/G$
is the
universal surface density
given by Eq. (\ref{lm22}), we get
\begin{equation}
S\sim \frac{k_B  c \sqrt{\Lambda}R^2}{Gm_{\Lambda}}.
\label{bh15}
\end{equation}
Interestingly, the entropy given by Eq. (\ref{bh15}) scales like the surface
$R^2$ of the object, similarly to the black hole entropy
(\ref{bh1}).\footnote{Inversely, a manner to understand why the surface
density of the dark matter halos has a universal value is to argue that their
entropy
given by Eq. (\ref{bh14}) should scale like $R^2$ (see Appendix
\ref{sec_post}).} This may
be connected to a form of holographic principle \cite{bousso}. Matching the
formulae (\ref{bh1}) and (\ref{bh15}), we find that
$m_{\Lambda}$ corresponds to the mass given by Eq. (\ref{bh7}). Inversely, if we
assume from the start that the logotropic dark fluid is composed of particles of
that mass (cosmons), we find that the entropy of dark matter halos coincides
with the
entropy of black holes.\footnote{Of course, we are not claiming that dark matter
halos
are black holes since they obviously do not fulfill the Schwarzschild relation
(\ref{bh2}). However, they may have the same entropy as black holes expressed
in terms of $R$ [see Eq. (\ref{bh1})].} On the other hand, since the surface
density of the Universe is
of the same order as the surface density of dark matter halos, the previous
formulae also
apply to the Universe as a whole and return the results of Appendix
\ref{sec_abhb}. This may be a form of justification, for reasons of
self-consistency, of Eq. (\ref{bh14}).

{\it Remark:} If we, alternatively,  define the entropy of dark matter halos by
$S\sim
k_B M/m_e$ where
$m_e$ is the electron mass and use $M\sim \Sigma_0 R^2$ with $\Sigma_0\sim
\Sigma_e$ where $\Sigma_e$ is the surface density of the electron given by Eq.
(\ref{lm27}), we obtain 
\begin{equation}
S\sim k_B \frac{R^2}{r_e^2},
\label{bh16}
\end{equation}
which is similar to the black hole entropy formula (\ref{bh1}) where the Planck
length $l_P$ is replaced by the classical radius of the electron $r_e$. It is
not clear, however, if this formula is physically relevant.

\subsection{Postulates: entropic principles}
\label{sec_post}

We can find a form of explanation of the different relations found in this
paper by making the following two postulates.

{\it Postulate 1:} We postulate that the entropy of the electron, the entropy of
dark matter halos and the entropy of the Universe (and possibly other
objects) is given by
\begin{equation}
S\sim \frac{k_B c^3R^2}{G\hbar},
\label{post1}
\end{equation} 
like the Bekenstein-Hawking entropy of black holes, where  $R$ is the radius of
the corresponding object.
This may be connected to a form of holographic principle stating
that the entropy is proportional to the area (instead of the volume). Therefore,
\begin{equation}
S_e/k_B\sim  \frac{c^3r_e^2}{G\hbar}   \qquad ({\rm electron})
\label{post2}
\end{equation} 
\begin{equation}
S/k_B \sim  \frac{c^3r_h^2}{G\hbar}\qquad ({\rm dark\,\, matter})
\label{post3}
\end{equation} 
\begin{equation}
S_{\Lambda}/k_B \sim  \frac{c^3R_{\Lambda}^2}{G\hbar}\qquad ({\rm
Universe})
\label{post4}
\end{equation} 

{\it Postulate 2:} We postulate that the entropy of the electron, the entropy of
dark matter halos and the entropy of the Universe (and possibly other
objects) is also  given by\footnote{Note
that this relation is {\it not} satisfied by black holes since $M_{\rm
BH}\propto R$
while $S_{\rm BH}\propto R^2$.}
\begin{equation}
S\sim k_B \frac{M}{m_{\Lambda}}, 
\label{post5}
\end{equation} 
where $M$ is the mass of the corresponding object and $m_{\Lambda}$ is the mass
defined by Eq. (\ref{bh7}). Therefore,
\begin{equation}
S_e/k_B\sim \frac{m_e}{m_{\Lambda}}\sim
\frac{\Sigma_{e}r_{e}^2}{m_{\Lambda}}\sim 10^{39}\qquad ({\rm
electron})
\label{post6}
\end{equation} 
\begin{equation}
S/k_B\sim \frac{M_h}{m_{\Lambda}}\sim
\frac{\Sigma_{0}r_h^2}{m_{\Lambda}}\qquad ({\rm dark \,\, matter})
\label{post7}
\end{equation} 
\begin{equation}
S_{\Lambda}/k_B\sim \frac{M_{\Lambda}}{m_{\Lambda}}\sim
\frac{\Sigma_{\Lambda}R_{\Lambda}^2}{m_{\Lambda}}\sim 10^{123}\qquad ({\rm
Universe})
\label{post8}
\end{equation} 

The comparison of Eqs. (\ref{post1}) and (\ref{post5}) directly implies that the
surface density of
the electron, the surface density of {\it all} the dark matter halos,  and the
surface
density of the Universe is (approximately) the same and has the typical value 
\begin{equation}
\Sigma\sim \frac{M}{R^2}\sim \frac{m_{\Lambda}c^3}{G\hbar}\sim
\frac{m_{\Lambda}}{M_P}\Sigma_P\sim \frac{c\sqrt{\Lambda}}{G},
\label{post9}
\end{equation} 
where $\Sigma_P=(c^7/\hbar G^3)^{1/2}=8.33\times 10^{64}\, {\rm g\, m^{-2}}$ is
the Planck surface density. Then, comparing this universal value with the
surface density of the electron [see Eq. (\ref{lm27})], we obtain the Weinberg
relation
\begin{equation}
\Lambda\sim \frac{m_e^6G^2c^6}{e^8}.
\label{post10}
\end{equation}

{\it Remark:} we have introduced the entropy of an electron [see Eqs.
(\ref{post2}) and (\ref{post6})] by analogy with the black hole entropy. If
these ideas are physically relevant, a notion of thermodynamics for the electron
(assuming
that it is made of $10^{39}$ subparticles of mass $m_{\Lambda}$) should be
developed. Again, the analogy with black holes (although, of course, an electron
is not a black hole) might be useful. 

\section{Large numbers and coincidences}
\label{sec_ln}

The ratio between the electric radius of the electron
$r_e=e^2/m_ec^2$ and its gravitational radius $r_g=2Gm_e/c^2$
is of the order of $e^2/Gm_e^2=4.17\times 10^{42}$. This
dimensionless number was computed by Weyl in 1919 \cite{barrow}. He was the
first to notice the presence of large dimensionless numbers in Nature. This led
Eddington \cite{eddington} and others to try to relate such large numbers to
cosmological quantities. In particular,  Eddington evaluated the total number of
particles in the Universe and found $N\sim 10^{79}$. He then tried to relate the
basic interaction strenghts and elementary particle masses to
this number. For example, it was observed by different authors that the
following quantities are of
the same order of magnitude (see Ref. \cite{barrow}, P. 224-231):
\begin{equation}
\frac{m_ec^2}{\hbar H_0}\sim
\frac{e^2}{Gm_e^2}\sim \sqrt{N}\sim
\left (\frac{M_P}{m_e}\right )^2\sim 10^{40}.
\label{ln1}
\end{equation}

These coincidences can
be easily understood from our results (\ref{lm32}) and (\ref{lm33}) which
express the mass and the charge of the electron in terms of
the cosmological constant. In order to avoid too much
digression, we shall replace the  Eddington number by\footnote{The
Eddington number corresponds typically to the number of protons in the
Universe, $N\sim M_{\Lambda}/m_p$, where $m_p$ is the proton mass. This number
was introduced before dark matter and dark energy were discovered. If the dark
fluid is made of cosmons of mass $m_{\Lambda}$, the number of particles in the
Universe is $N_{\Lambda}= M_{\Lambda}/m_{\Lambda}\sim 10^{123}$ giving another
interpretation to the famous number $123$. This number should supersede
the Eddington number.}
\begin{equation}
N_e=\frac{M_{\Lambda}}{m_e}\sim e^{2/(3B)}\sim 10^{80}.
\label{ln2}
\end{equation}
On the other hand, combining our
results, we find
\begin{equation}
\frac{m_e c^2}{\hbar H_0}\sim \frac{m_e}{m_{\Lambda}}\sim
e^{1/(3B)}\sim  10^{40},
\label{ln3}
\end{equation}
\begin{equation}
\frac{e^2}{Gm_e^2}\sim e^{1/(3B)}\sim 10^{40},
\label{ln4}
\end{equation}
\begin{eqnarray}
\frac{M_P}{m_e}\sim e^{1/(6B)}\sim 10^{20},
\label{ln5}
\end{eqnarray}
leading to the equivalents from Eq. (\ref{ln1}). We also note that
\begin{eqnarray}
\frac{1}{m_e^4}\left (\frac{\hbar c}{G}\right )^2\sim \left
(\frac{M_P}{m_e}\right )^4\sim 10^{80}\sim N_e,
\label{ln6}
\end{eqnarray}
which is one of the ``coincidences'' pointed out by Chandrasekhar
\cite{chandranature}.

In a sense, these results arise from the Weinberg relation (\ref{w4}) that
has been found by different authors (see footnote 7).
Nevertheless we believe
that our approach is original and may bring new light on the subject. In
particular, we have proposed a form of common explanation of these different
``coincidences'' in terms of entropic principles (see Appendix \ref{sec_post}).

\section{Thermodynamics of the logotropic dark fluid}
\label{sec_logt}

Let us try to relate the results of the previous Appendices to the
thermodynamics of the logotropic dark fluid. 

We assume that the Universe is
filled with a dark fluid at temperature $T$. From
the first principle of thermodynamics, one can derive the thermodynamic equation
\cite{weinbergbook}:
\begin{equation}
\label{t1}
\frac{dP}{dT}=\frac{1}{T}(\epsilon+P).
\end{equation}
If the dark fluid is described by a barotropic equation of state of the form 
$P=P(\epsilon)$, Eq. (\ref{t1}) can be integrated to obtain the relation
$T=T(\epsilon)$ between the temperature and the energy density. On the other
hand, the entropy of the dark fluid in a volume $a^3$ is given by
\cite{weinbergbook}:
\begin{equation}
\label{t2}
S=\frac{a^3}{T}(P+\epsilon).
\end{equation}
From the Friedmann equations, one can show
 that the entropy of the Universe is conserved:
${dS}/{dt}=0$ \cite{weinbergbook}.

The previous results are general. Let us now apply them to the logotropic dark
fluid. According to Eqs. (\ref{lm5}) and (\ref{lm6}), the equation of
state $P=P(\epsilon)$ of the logotropic dark fluid is given by the reciprocal of
\cite{lettre,jcap}:
\begin{equation}
\label{t3}
\epsilon=\rho_P e^{P/A}c^2-P-A.
\end{equation}
Eq. (\ref{t1}) with Eq. (\ref{t3}) is easily integrated giving
\begin{equation}
\label{t4}
T=\frac{\rho_Pc^2}{K}\left (1-\frac{A}{\rho c^2}\right ),
\end{equation}
where $K$ is a constant of integration and we have used Eq. (\ref{lm5}).
Substituting Eqs. (\ref{t3}) and (\ref{t4}) into Eq. (\ref{t2}), and using Eqs.
(\ref{lm3}) and (\ref{lm5}), we find that
\begin{equation}
\label{t5}
S=K\frac{\rho_0}{\rho_P}.
\end{equation}
We explicitly check on this expression that the entropy of the Universe is
conserved.
Furthermore, since the entropy is positive, we must have $K>0$. Considering Eq.
(\ref{t4}), we note
that the temperature is positive when $\rho>\rho_M=A/c^2$ and negative when
$\rho<\rho_M=A/c^2$, that is to say when the Universe becomes phantom
\cite{epjp,lettre}.\footnote{This is a general result \cite{cosmopoly3} which
can be obtained from
Eq. (\ref{t2}) using the fact that the entropy is constant and positive. We see
that the sign of the
temperature coincides with the sign of $P+\epsilon$. As a result, the
temperature is positive in a normal Universe ($P>-\epsilon$) and negative in a
phantom Universe ($P<-\epsilon$).}

We can  determine the constant $K$ by assuming that the
entropy of the logotropic dark fluid is given by 
\begin{equation}
\label{t6}
S\sim k_{B}\frac{M_{\Lambda}}{m_{\Lambda}}\sim  10^{123}\, k_B
\end{equation}
as in Appendix \ref{sec_abh}. Noting that the ``true'' entropy is
obtained by multiplying Eq. (\ref{t2}) by $R_{\Lambda}^3$ (since we have taken
$a=1$ at the present time), and comparing
Eqs. (\ref{t5}) and
(\ref{t6}), we obtain 
\begin{equation}
\label{t7}
K\sim k_B \frac{\rho_P}{m_{\Lambda}}.
\end{equation}
As a result, the temperature of the logotropic dark fluid is given by
\begin{equation}
\label{t8}
k_B T\sim m_{\Lambda}c^2\left (1-\frac{B\rho_{\Lambda}}{\rho}\right ),
\end{equation}
where we have used Eq. (\ref{lm12}). In the ``early'' Universe
$\rho\gg\rho_{\Lambda}$ we find  that\footnote{We recall
that the logotropic model, which is a
unification of dark matter and dark energy, is not valid in the very early
Universe
corresponding to the big-bang, the inflation era, and the radiation era.
Therefore, the temperature $m_{\Lambda}c^2$ corresponds to the temperature
of the dark fluid in the matter era, i.e., when the rest-mass energy of the dark
fluid overcomes its internal energy (see Sec. \ref{sec_lm}). We emphasize that
the
temperature $T$ of the logotropic dark fluid is different from the temperature
of radiation and of any other standard temperature.  We also note that the
corresponding temperature in the $\Lambda$CDM model is not defined since Eq.
(\ref{t1}) breaks down when $P=0$.} 
\begin{equation}
T\simeq 
m_{\Lambda}c^2/k_B=2.41\times 10^{-29}\, {\rm K}.
\end{equation}
In the late Universe $\rho\ll\rho_{\Lambda}$
we
find that 
\begin{equation}
k_B T\sim -m_{\Lambda}c^2B\rho_{\Lambda}/\rho\propto -a^3.
\end{equation}

{\it Remark:} In Ref. \cite{jcap} we have shown that the logotropic constant $B$
could be interpreted as a dimensionless logotropic temperature 
\begin{equation}
\label{t9}
B=\frac{k_B T_{\rm L}}{m_{\Lambda}c^2}
\end{equation}
in a generalized thermodynamical framework \cite{epjp,lettre}. This shows that
at least two
temperatures exist for the logotropic dark fluid, a time-varying temperature
$T$ and a constant temperature $T_{\rm L}$. They become equal when
\begin{equation}
\label{t10}
\frac{\rho_*}{\rho_{\Lambda}}\sim \frac{B}{1-B}\sim 3.54\times 10^{-3},
\end{equation}
\begin{equation}
\label{t11}
a_*\sim \left (\frac{\Omega_{\rm m,0}}{\Omega_{\rm de,0}}\frac{1-B}{B}\right
)^{1/3}\sim 5.01.
\end{equation}

\section{The mass of the bosonic dark matter particle}
\label{sec_mdm}

It has been suggested that dark matter may be made of
bosons
(like ultralight axions) in the form of Bose-Einstein condensates
(BECs) \footnote{See, e.g., the bibliography of Ref.
\cite{abriljeans}
for an
exhaustive list of references. The possible connections between the BECDM
model and the
logotropic model
will be investigated in a future paper \cite{forthcoming}.} We can use
the results of the present  paper to predict the mass $m$ of the bosonic dark
matter
particle in terms of the cosmological constant $\Lambda$. We assume that the
smallest and most compact dark matter
halo that is observed corresponds to the
ground state of a
self-gravitating BEC (to fix the ideas we assume that this halo is the dSphs
Fornax with a mass
$M\sim 10^8\, M_{\odot}$ and a radius $R\sim 1\, {\rm kpc}$). For noninteracting
bosons, it can be shown by solving
the Gross-Pitaevskii-Poisson equations (see, e.g., Sec. III.B.1
of
\cite{becmodel}) that the
mass $(M_h)_{\rm min}$, the radius $(r_h)_{\rm min}$ and the central density
$(\rho_0)_{\rm max}$ of this ultracompact halo (ground state)
are related to each other by the relations
\begin{equation}
M_h=1.91\, \rho_0 r_h^3 \qquad  {\rm and} \qquad M_h
r_h=1.85\, \frac{\hbar^2}{Gm^2}. 
\label{add}
\end{equation}
As a result, its surface density is given
by
\begin{equation}
\Sigma_0=0.153\, \frac{G^2m^4M_h^3}{\hbar^4}.
\label{mdm1}
\end{equation}
On the other hand, the minimum mass of dark matter halos may be obtained from a
quantum Jeans instability theory (see, e.g., Ref. \cite{abriljeans}) giving the
result
\begin{equation}
M_J=\frac{1}{6}\pi\left
(\frac{\pi^3\hbar^2\rho_{\rm dm,0}^{1/3}}{Gm^2}\right
)^{3/4}.
\label{mdm2}
\end{equation}
For usually considered values of the boson mass, of the order of $m\sim
10^{-22}\, {\rm eV/c^2}$, the Jeans mass  $M_J \sim 10^7\,
M_{\odot}$ is indeed of the order of the minimum mass $(M_h)_{\rm min} \sim
10^8\, M_{\odot}$ of observed dark matter halos.
There may be, however, a numerical factor of order $10$ between $M_J$ and
$(M_h)_{\rm min}$.
For that reason, we introduce a prefactor $\chi$ and write $(M_h)_{\rm min}=\chi
M_J$. Using $\rho_{\rm dm,0}=({\Omega_{\rm dm,0}}/{\Omega_{\rm
de,0}})\rho_{\Lambda}=({\Omega_{\rm dm,0}}/{\Omega_{\rm de,0}})
({\Lambda}/{8 \pi G})$, we get
\begin{equation}
(M_h)_{\rm min}=\chi \frac{\pi^3}{6}\left (\frac{\Omega_{\rm
dm,0}}{8\Omega_{\rm
de,0}}\right )^{1/4}\frac{\hbar^{3/2}\Lambda^{1/4}}{Gm^{3/2}}.
\label{mdm3}
\end{equation}
Then, using Eq. (\ref{mdm1}), we obtain
\begin{equation}
\Sigma_0= 0.153\,\chi^3 \frac{\pi^9}{216}\left (\frac{\Omega_{\rm
dm,0}}{8\Omega_{\rm de,0}}\right
)^{3/4}\frac{\hbar^{1/2}\Lambda^{3/4}}{Gm^{1/2}}.
\label{mdm4}
\end{equation}
Comparing this expression with Eq. (\ref{lm22}), we predict  that the mass of
the bosonic particle is given by
\begin{equation}
m=\chi^6\frac{0.0234\pi^{20}}{1458B\xi_h^2}\left
(\frac{\Omega_{\rm
dm,0}}{8\Omega_{\rm de,0}}\right
)^{3/2}\frac{\hbar\sqrt{\Lambda}}{c^2}=15397\chi^6\frac{
\hbar\sqrt{\Lambda}}{c^2}.
\label{mdm5}
\end{equation}
We see that the mass of the bosonic dark matter particle is equal to the mass
scale
$m_{\Lambda}\sim 10^{-33}\,
{\rm eV/c^2}$ given by Eq. (\ref{bh7}) multiplied by a huge numerical factor of
order $10^{10}$ (for $\chi\sim 10$). This gives $m\sim 
10^{-23}\,
{\rm eV/c^2}$ which is the correct
order
of magnitude of the mass of ultralight axions usually considered
\cite{marshrevue}. We note
that this result has been obtained independently from the observations, except
for the value of $\Lambda$ and the other fundamental
constants (Planck scales). 

\eject

\end{document}